\documentclass[modern]{aastex62}
\usepackage{natbib}

\newcommand{\source}{J1706}

\shorttitle{The 38 Minute Orbit of IGR J17062$-$6143}
\shortauthors{Strohmayer et al.}

\begin{document}

\title{NICER Discovers the Ultracompact Orbit of the Accreting
  Millisecond Pulsar IGR J17062$-$6143}

\author{T. E. Strohmayer} 
\affil{Astrophysics Science Division and Joint Space-Science Institute,
  NASA's Goddard Space Flight Center, Greenbelt, MD 20771, USA}

\author{Z. Arzoumanian} 
\affil{X-ray Astrophysics Laboratory,
  Astrophysics Science Division, NASA's Goddard Space Flight Center,
  Greenbelt, MD 20771, USA} 

\author{S. Bogdanov} 
\affil{Columbia Astrophysics Laboratory, Columbia University, 550 
West 120th Street, New York, NY 10027, USA}

\author{P. M. Bult}
\affil{Astrophysics Science Division,
  NASA's Goddard Space Flight Center, Greenbelt, MD 20771, USA}

\author{D. Chakrabarty} 
\affil{MIT Kavli Institute for Astrophysics
  and Space Research, Massachusetts Institute of Technology,
  Cambridge, MA 02139, USA}

\author{T. Enoto} 
\affil{The Hakubi Center for Advanced Research and
  Department of Astronomy, Kyoto University, Kyoto 606-8302, Japan}

\author{K. C. Gendreau}
\affil{X-ray Astrophysics Laboratory,
  Astrophysics Science Division, NASA's Goddard Space Flight Center,
  Greenbelt, MD 20771, USA} 

\author{S. Guillot} 
\affil{CNRS, IRAP, 9 avenue du Colonel Roche, BP
  44346, F-31028 Toulouse Cedex 4, France} 
\affil{Université de
  Toulouse, CNES, UPS-OMP, F-31028 Toulouse, France}

\author{A. K. Harding}
\affil{Astrophysics Science Division and Joint Space-Science Institute,
  NASA's Goddard Space Flight Center, Greenbelt, MD 20771, USA}

\author{W. C. G. Ho} 
\affil{Department of Physics and Astronomy,
  Haverford College, 370 Lancaster Ave., Haverford, PA 19041, USA}
\affil{Mathematical Sciences, Physics and Astronomy, and STAG Research
  Centre, University of Southampton, Southampton SO17 1BJ, United
  Kingdom}

\author{J. Homan} 
\affil{Eureka Scientific, Inc., 2452 Delmer Street,
  Oakland, CA 94602, USA}
\affil{SRON, Netherlands Institute for Space Research, 
Sorbonnelaan 2, 3584 CA Utrecht, The Netherlands}

\author{G. K. Jaisawal}
\affil{National Space Institute, Technical University of Denmark, 
Elektrovej 327-328, DK-2800, Lyngby, Denmark}

\author{L. Keek} 
\affil{Department of Astronomy, University of Maryland College Park, 
MD 20742, USA}

\author{M. Kerr}
\affil{Space Science Division, Naval Research Laboratory, Washington, 
DC 20375-5352, USA}

\author{S. Mahmoodifar} 
\affil{Astrophysics Science Division and Joint
  Space-Science Institute, NASA's Goddard Space Flight Center,
  Greenbelt, MD 20771, USA}

\author{C. B. Markwardt}
\affil{X-ray Astrophysics Laboratory,
  Astrophysics Science Division, NASA's Goddard Space Flight Center,
  Greenbelt, MD 20771, USA} 

\author{S. M. Ransom}
\affil{National Radio Astronomy Observatory, Charlottesville, VA 22903, USA}

\author{P. S. Ray} 
\affil{Space Science Division, Naval Research Laboratory, Washington, 
DC 20375-5352, USA}

\author{R. Remillard}
\affil{MIT Kavli Institute for Astrophysics
  and Space Research, Massachusetts Institute of Technology,
  Cambridge, MA 02139, USA}

\author{M. T. Wolff}
\affil{Space Science Division, Naval Research Laboratory, Washington, 
DC 20375-5352, USA}

%\collaboration{(and other NICER authors)}

\begin{abstract}

We present results of recent {\it Neutron Star Interior Composition
  Explorer} observations of the accreting millisecond X-ray pulsar IGR
J17062$-$6143 that show that it resides in a circular, ultracompact
binary with a 38 minute orbital period.  {\it NICER} observed the
source for $\approx 26$ ksec over a 5.3 day span in 2017 August, and
again for 14 and 11 ksec in 2017 October and November, respectively. A
power spectral analysis of the August exposure confirms the previous
detection of pulsations at 163.656 Hz in {\it Rossi X-ray Timing
  Explorer} data, and reveals phase modulation due to orbital motion
of the neutron star. A coherent search for the orbital solution using
the $Z^2$ method finds a best-fitting circular orbit with a period of
2278.21 s (37.97 min), a projected semi-major axis of 0.00390 lt-sec,
and a barycentric pulsar frequency of 163.6561105 Hz. This is
currently the shortest known orbital period for an AMXP. The mass
function is $9.12 \times 10^{-8}$ $M_{\odot}$, presently the smallest
known for a stellar binary. The minimum donor mass ranges from
$\approx 0.005 - 0.007$ $M_{\odot}$, for a neutron star mass from 1.2
- 2 $M_{\odot}$. Assuming mass transfer is driven by gravitational
radiation, we find donor mass and binary inclination bounds of $0.0175
- 0.0155 M_{\odot}$ and $ 19^{\circ} < i < 27.5^{\circ}$, where the
lower and upper bounds correspond to 1.4 and 2 $M_{\odot}$ neutron
stars, respectively. Folding the data accounting for the orbital
modulation reveals a sinusoidal profile with fractional amplitude
$2.04 \pm 0.11 \%$ (0.3 - 3.2 keV).
%We discuss the implications of our findings the object's nature.

\end{abstract} % Maximum 250 words for ApJ Letter
\keywords{stars: neutron --- stars: oscillations --- stars: rotation ---
X-rays: binaries --- X-rays: individual (IGR J17062$-$6143) --- methods:
data analysis}

\section{Introduction}

\label{sec:introduction} 

The accreting neutron star binary IGR J17062$-$6143 (hereafter, J1706)
is one of the most recently identified accreting millisecond X-ray
pulsars (AMXP).  First observed in outburst in 2006
\citep{2007A&A...467..529C, 2008ATel.1840....1R, 2008ATel.1853....1R},
it has since then been persistently accreting at luminosities in the
range 5.8 - $7.5 \times 10^{35}$ erg s$^{-1}$ (2 - 20 keV), assuming a
distance of 7.3 kpc \citep{2013ApJ...767L..37D, 2017ApJ...836..111K,
  2017ApJ...836L..23S, 2018MNRAS.475.2027V}. The object's neutron star
nature was first revealed by the detection of thermonuclear X-ray
bursts.  The first of these was observed by {\it Swift} in 2012
\citep{2013ApJ...767L..37D}, and most recently
\citet{2017ApJ...836..111K} reported on {\it Swift} observations of a
long duration burst first detected with {\it MAXI}
\citep{2015ATel.8241....1N} that was likely powered by burning of a
deep helium layer. The properties of these long duration (tens of
minutes) thermonuclear X-ray bursts are consistent with the
accumulation of helium rich material on the neutron star surface,
which could be accommodated by accretion from a degenerate helium
dwarf in an ultracompact system.  However, accretion of hydrogen-rich
fuel under certain conditions can also lead to thick, combustible
helium layers, so the observation of apparently helium-powered nuclear
flashes is not necessarily a definitive indication of an ultracompact
system \citep{1981ApJ...247..267F,
  2006ApJ...652..559G}. \citet{2017ApJ...836L..23S}, hereafter SK17,
reported the detection ($4.3 \sigma$) of 163.656 Hz pulsations in a
single $\approx 1200$ s observation with the {\it Rossi X-ray Timing
  Explorer} ({\it RXTE}; \cite{1993A&AS...97..355B}). They found a
fractional pulsed amplitude (after background subtraction) of $9.4 \pm
1.1 \%$, but could not determine the orbital period of the system due
to the single, short {\it RXTE} observation. They were able to place a
lower limit on the orbital period of about 17 minutes.

The source has recently been studied extensively with {\it Swift},
{\it NuSTAR}, {\it Chandra} and {\it XMM-Newton}. For example,
\citet{2017MNRAS.464..398D} reported the presence of Fe K$\alpha$
reflection features in {\it NuSTAR} data, modeling of which suggested
an inner disk that may be truncated out to $\approx 100 R_g$, where
$R_g = GM/c^2$.  Most recently, \citet{2018MNRAS.475.2027V} presents
results of simultaneous {\it NuSTAR} and {\it XMM-Newton}
observations. They report the presence of reflection features as well,
and suggest a similarly truncated disk as in
\citet{2017MNRAS.464..398D}.  They note, however, that a disk
extending down to the neutron star cannot be excluded if the binary
inclination is very low.  Based on analysis of {\it XMM-Newton}
Reflection Grating Spectrometer (RGS) data they also suggest the
system may have an oxygen-rich circumbinary environment, perhaps due
to an outflow.  Interestingly, they also searched for pulsations using
the {\it XMM-Newton} EPIC timing mode data but did not detect them.
They placed an upper limit on the pulsed fraction in those data of
$5.4 \%$ ($0.5 - 10$ keV). They concluded that the persistently faint
X-ray luminosity could be indicative of either an ultracompact binary
system or perhaps magnetic truncation, but the spectroscopic data
alone were not decisive between these two
possibilities. \citet{2018arXiv180103006H} have recently reported on
broad-band optical to near-infrared (NIR) photometry of \source{} that
they modeled as emission from an irradiated accretion disk. Their
modeling indicates an accretion disk size consistent with an
ultracompact orbit, and they argued for an orbital period in the range
from 0.4 - 1 hr. Additionally, their optical spectroscopy showed no
H-$\alpha$ emission, consistent with a hydrogen-deficient donor and an
ultracompact system. Thus, sensitive, new timing observations to
determine the binary orbital parameters, and the nature of the system,
were clearly warranted.

In this paper we report results of recent {\it Neutron Star Interior
  Composition Explorer} (NICER) observations of \source{}. The
principal goals of the {\it NICER} observing campaign were to confirm
(or not) the {\it RXTE} detection of 163.656 Hz pulsations and, if
pulsations could be detected, to determine the system's orbital
parameters. We show below that the new {\it NICER} data confirm that
\source{} is an 163.656 Hz pulsar, and also reveal an ultracompact
orbit with similarities to other ultracompact AMXPs
\citep{2012arXiv1206.2727P}.  The plan of the paper is as follows. We
first describe the observations, data selection, and our initial
pulsation search and detection, confirming that \source{} is a 163.656
Hz pulsar.  We next discuss our orbit search and detection, and we
summarize the properties of the system given the orbit solution.  We
conclude with a brief summary and discussion of the implications of
our findings for the nature of \source{}.

%We first present the results of our pulsation search of \source{} and
%describe the discovery of pulsations (Section~\ref{sec:pulsation}). We
%(Section~\ref{sec:discussion}). 
%We conclude that  . (Section~\ref{sec:conclusions}).

\section{NICER Observations and Pulsation Search}

\label{sec:observations} 
{\it NICER} was installed on the International Space Station ({\it
  ISS}) in 2017 June, and began full science operations after a one
month checkout and verification period.  {\it NICER} is optimized for
low background, high throughput, fast timing observations across the
$0.2 - 12$ keV band \citep{2012SPIE.8443E..13G}, achieving an absolute
timing precision of $\approx 100$ ns with the aid of a GPS
receiver. We obtained with {\it NICER} $26$ ks of good exposure on
\source{} in the time window spanning 2017 August 9 - 15.  Additional
observations were obtained in October and November, but we focused on
the August data for our initial pulsation search.  We processed and
analyzed the data using HEASOFT version 6.22 and NICERDAS
2017-09-06\_V002.  We barycentered the data using the tool {\it
  barycorr} employing the {\it DE200} solar system ephemeris and
source coordinates {\it R.A.} $=256.5677 ^{\circ}$, decl. $=-61.7113
^{\circ}$ \citep{2008ATel.1840....1R}.  After data processing and
selection we identified 58 good time intervals (GTIs) of at least 50 s
duration in the August data, for a total of $26$ ks of on-source
exposure.  The on-source dwell times with {\it NICER} tend to be
somewhat shorter than for free-flying, low-Earth orbit observatories.
Figure 1 shows the resulting light curve accumulated in 16 s bins, and
including events with energies in the range from 0.3 to 5 keV. The
average count rate was $\approx 31$ s$^{-1}$, which is consistent with
the expected rate estimated using source flux and spectra from recent
observations \citep{2017MNRAS.464..398D, 2017ApJ...836..111K}.  We do
see evidence for variations in the background counting rates,
particularly in the August data. This is most evident in the band
above $\approx 5$ keV.  Based on the average count rate spectrum we
estimate that the {\it NICER} background in the 0.3 to 5 keV band is,
on average, less than the source count rate by a factor of $\approx
15$. We note that we did not observe any thermonuclear X-ray bursts
from the source, but given the long recurrence time, this is not very
surprising \citep{2017ApJ...836..111K}.

For our pulsation search we further limited the upper end of the
energy band to 3.2 keV due to the higher backgrounds present in some
dwells. Our choice here reflected a trade-off between either removing
a substantial number of dwells completely or reducing the upper energy
threshold somewhat, and thereby allowing us to utilize most of the
dwells.  To search for pulsations we computed a light curve (0.3 - 3.2
keV) of the full August dataset sampled at 4096 Hz.  This light curve
spans 500 ksec ($\approx 5.8$ days) and has $(5\times 10^5)\times 4096
= 2.048\times 10^9$ bins.  We computed a FFT power spectrum of this
light curve and searched in the frequency range in which pulsations
were reported by SK17.  Figure 2 shows the resulting power spectrum,
normalized as in \cite{1983ApJ...266..160L}, in the vicinity of the
163.656 Hz pulse frequency.  The red, vertical dashed lines denote the
approximate range of pulse frequency detected in the {\it RXTE}
observations.  An excess of signal power consistent with this
frequency range is clearly evident.  The highest power in the plotted
frequency range has a value of 56.3.  The expected noise power
distribution is a $\chi^2$ distribution with 2 degrees of freedom.
The probability to exceed this value (56.3) in a single trial is
$6\times 10^{-13}$. There are 15,000 frequency bins in the range from
163.64 to 163.67 Hz. This gives a chance occurrence probability of $9
\times 10^{-9}$ for the highest observed power only. Since additional
excess powers are present as well, this is an extremely conservative
significance estimate.

In Figure 2 one can see that the pulsar signal is comprised of two
main sidebands, each of which is modulated by a number of finely
spaced peaks.  The first, most significant, sideband is that near the
center of the frequency band denoted by the red, vertical lines, and
the second is near the high end of the band. There is likely a third
sideband mid-way between these two, but it is weak enough that it is
harder to discern above the noise.  The presence of such sidebands in
the power spectrum is a strong indication of the presence of phase
modulation due to orbital motion of the pulsar
\citep{2003ApJ...589..911R}.  We do not detect any excess power at
higher harmonics of the pulsar frequency. We note that the finely
spaced peaks modulating each sideband are consistent with the {\it
  ISS} orbit period, thus these result from the incomplete sampling
(gaps) in the time series, that is, {\it NICER}'s window function.

As a further test we also computed power spectra of some of the long
individual on-source dwells, and then averaged these.  We used 21 of
the longest dwells with exposures ranging from 1294 s (longest) to 545
s (shortest). These intervals are indicated in Figure 1 by vertical
dashed lines drawn at the center of each interval. For each of these
intervals we computed a light curve sampled at 4096 Hz and with a
duration of 2048 s.  Since this is longer than each of the individual
exposures we padded the light curves to 2048 s using the mean value
determined from the good exposure in each dwell. This procedure
ensures that the same Fourier frequency spacing is used for each
interval and facilitates simple averaging of the resulting power
spectra.  We then computed FFT power spectra for each dwell and
averaged them. The resulting power spectrum is shown in Figure 3, and
clearly shows an excess of power at the same frequency as is evident
in Figure 2. Indeed, we see the same basic signal structure of two
dominant sidebands. Thus, {\it NICER} clearly detects pulsations from
\source{} in a frequency range consistent with the earlier {\it RXTE}
detection, and moreover, the sidebands in the power spectra are
strongly indicative of accelerated motion of the pulsar.

\section{Searching for the Orbit}

Having recovered pulsations from \source{} with {\it NICER} we next
began a search for the orbital parameters.  The combination of a weak
pulsed signal and relatively short uninterrupted exposures means that
it is not really possible to closely track the pulse frequency
variations with time around the orbit, particularly if the orbital
period is short compared to the typical gap in exposure.  It is
therefore not possible to directly ``see'' the orbital frequency
evolution with time in, for example, a dynamic power spectrum.  This
makes it more challenging to deduce the orbit.  Nevertheless, orbital
motion of the pulsar introduces periodic light travel time
delays/advances that depend on its orbital phase.  As noted above,
these produce a characteristic sideband structure in power spectra
computed from a light curve that samples at least several orbital
periods \citep{2003ApJ...589..911R}.  In principle, one can measure
the orbital period by detecting this sideband structure in the power
spectrum, as the frequency spacing of the phase modulation sidebands
is set by the orbital period. However, the complex window function
(due to the data gaps) associated with {\it NICER}'s observing windows
makes it challenging in the present case to directly infer
the orbital period in this way.

%The amplitude of the phase modulation, $A_{\phi}$, is essentially the
%light travel time across the pulsar's projected semi-major axis,
%$a_x\sin i/c$, expressed as a fraction of the pulsar spin period.  In
%the limit that the amplitude is small, the sideband structure
%simplifies to a dominant peak at the pulsar spin frequency and a pair
%of weaker, adjacent sideband peaks separated by $\Delta f =
%1/P_{orb}$.  The power spectrum of the full August dataset (Figure 2),
%with a small number of widely spaced, prominent sidebands, is somewhat
%suggestive of this, and would seem to imply a compact orbit ($P_{orb}
%< 1$ hr) with a relatively small phase modulation amplitude.

Because of the challenges outlined above we employed a coherent, grid
search for the orbital parameters using the $Z^2_n$ statistic
\citep{1983A&A...128..245B, 2002ApJ...577..337S}, and since there is
no evidence for harmonic signals we began by restricting the analysis
to $n=1$.  The observed population of AMXPs are all in highly circular
orbits, and indeed there are good theoretical arguments why this
should be the case, so we began our search with circular orbit models.
We also start with the assumption that the pulsar frequency does not
vary significantly across the August data epoch. We used the Blandford
\& Teukolsky (1976) relativistic orbit model to parameterize the time
delays, and with the assumption of circularity, we have a four
parameter search space; the pulsar frequency, $\nu_0$, projected
semi-major axis, $a_x \sin i/c$, orbital period, $P_{orb}$, and the
epoch of mean longitude equal to zero, $T_0$. Mean longitude is the
orbit phase angle measured from the ascending node. For a circular
orbit, the pulsar is ``behind'' the companion star at mean longitude
of $90^{\circ}$. We evaluate the statistic,
%The epoch of the ascending node is MJD 57974.82795462 (TDB)
\begin{equation}
Z_1^2 = \left ( \sum_{j=1}^N \cos\phi_j \right )^2  + 
  \left ( \sum_{j=1}^N \sin\phi_j \right )^2  \; ,
\end{equation}
where, $\phi_j = \nu_0(t_j + \Delta t_{BT} (t_j, a_x\sin i/c, P_{orb},
T_0))$, $\Delta t_{BT}$ is the binary time delay model
\citep{1976ApJ...205..580B} as a function of orbital parameters, and
$t_j$ are the barycentric photon arrival times.  Since such ``brute
force'' searches can be computationally expensive, we began by
searching a subset of the full August dataset.  For this we used the
more densely sampled portion of the light curve, the portion of Figure
1 between about 1.5 and 2.2 days. We set up grids of values spanning
the relevant ranges for each parameter. We used the recovered signal
in the power spectra (Figures 2 and 3), as well as prior results to
guide these choices. For example, SK17 used the {\it RXTE} data to
place a lower limit on the orbital period of about 17 min. As noted
above, the sideband structure in the power spectrum is suggestive of a
compact orbit. Based on this we confined our initial search to orbital
periods between 10 and 90 min. We used the power spectral results to
bound both $\nu_0$ and $a_x\sin i/c$. Finally, we employed a sampling
in $T_0$ equivalent to $2^{\circ}$ of orbital phase.  We then computed
$Z_1^2$ for all combinations of parameters to find candidate solutions
with large $Z_1^2$. This procedure yielded a candidate orbit solution
with $Z_1^2 = 77.1$, an orbital period of $P_{orb} = 2278$ s, $a_x\sin
i/c = 0.00393$ lt-sec, and $\nu_0 = 163.65611055$ Hz.  We did not find
any other comparable $Z^2_1$ maxima within the range of parameter
space searched.

Since this result was obtained from a subset of the August data, we
next attempted to coherently add all the additional August data
segments. We did this by adding data segments one at a time into the
total $Z_1^2$ sum and then used the {\it IDL} function
minimizer/maximizer {\it tnmin} to optimize the solution
\citep{2009ASPC..411..251M}.  In each case $Z_1^2$ increased in a
monotonic fashion, and the orbit parameters remained consistent.
Phasing up all the August data in this way resulted in a peak value of
$Z_1^2 = 196.1$. Recall that in the case of pure Poisson noise, the
$Z_1^2$ statistic is distributed as $\chi^2$ with 2 degrees of
freedom, and a value this high has a chance probability (single trial)
of $2.6 \times 10^{-43}$ (13.8 $\sigma$). The pulsed signal after
accounting for the orbital phase delays is dramatically stronger than
with no orbit correction, as expected. Figure 4 compares these two
signals. The curves show $Z_1^2$ evaluated on a grid of $\nu_0$ with
orbit phase delays included (black), and without (red). We note that,
as in Figure 2, the modulating ``comb'' of finely spaced sub-peaks
results from the observing window function. Using the full August data
set we find the best orbital solution has $P_{orb} = 2277.89 \pm 0.48
$ s, $a_x\sin i/c = 0.00395 \pm 0.0003$ lt-sec, $\nu_0 = 163.65611058
\pm 2.7 \times 10^{-7}$ Hz, and $T_0 =$ MJD $57974.82835 \pm 0.0007$
(TDB), where we quote nominal $3\sigma$ errors for a single parameter
by finding the values for each parameter at which $\Delta Z_1^2 = 9$
\citep{2002ApJ...575L..21M}.

We then carried out similar analyses on the October and November data
segments, treated separately, and found consistent results.  Finally,
we combined data across all epochs, and found a peak $Z_1^2 =
355.4$. We again determined confidence regions using $\Delta Z_1^2 =
9$, and found the best solution has $P_{orb} = 2278.208 \pm 0.012 $ s
(37.97 min), $a_x\sin i/c = 0.00389 \pm 0.0002$ lt-sec, $\nu_0 =
163.656110496 \pm 9 \times 10^{-9}$ Hz, and $T_0 =$ MJD $57974.82795
\pm 0.00028$ (TDB).  The timing solution is summarized in Table 1.

Next, we allowed the eccentricity, $e$, to be non-zero, but this did
not result in a significant increase in $Z_1^2$, and we placed an
upper limit on it of $e < 0.03$ ($1\sigma$, $\Delta Z_1^2 = 1$).
Using our best orbit solution we phase-folded all the data, and fit
the resulting pulse profile (0.3 - 3.2 keV) with a sinusoid, $A +
B\sin(\phi - \phi_0)$. The fit is excellent, with a minimum $\chi^2 =
8.6$ (13 degrees of freedom), and the implied fractional pulsed
amplitude is $B/A = 2.04 \pm 0.11 \%$. Figure 5 shows the resulting
pulse profile and fitted model. We did not detect any harmonic
signals.  We note that the pulsed amplitude measured with {\it NICER}
is comfortably below the upper limits reached in the recent pulsation
search with {\it XMM-Newton} reported by \citet{2018MNRAS.475.2027V},
which likely explains why they did not detect the pulsations. We also
computed pulse phase residuals using the best orbit solution. We show
these in Figure 6, where we have added in the orbit-predicted phase
delay in order to visually show the size of the delays.  We did not
find any statistically significant long-term trends in these
residuals.  Finally, we allowed for a constant pulsar spin frequency
derivative, $\dot\nu$, in the timing model, and recomputed $Z_1^2$ on
a grid of $\nu_0$ and $\dot\nu$ values. We found no significant
increase in $Z_1^2$, and we derived the following limits, $-6 \times
10^{-15} < \dot\nu < 4 \times 10^{-15}$ Hz s$^{-1}$ ($1\sigma$).

% table 1
\begin{table*}
\renewcommand{\arraystretch}{1.5}
\caption{Timing Parameters for IGR J17062-6143}
\scalebox{0.95}{
\begin{tabular}{lc}
\tableline\tableline
Parameter & Value \\
%\end{tabular}
\tableline
%\begin{tabular}{lc}
Right ascension, $\alpha$ (J2000) & $256.5677^{\circ}$ \\
Declination, $\delta$ (J2000) & $-61.7113^{\circ}$ \\
Barycentric pulse frequency, $\nu_0$ (Hz) & 163.656110049(9) \\
Pulsar frequency derivative, $\dot\nu$ (Hz s$^{-1}$) & $-6\times 10^{-15} < \dot\nu < 4 \times 10^{-15}$ \\
Projected semi-major axis, $a_x \sin(i) / c$ (lt - s) & 0.0039(2) \\
Binary orbital period, $P_{orb}$ (s) & 2278.208(12) \\
Time of mean longitude equal to zero, $T_0$ & MJD 57974.82795(28) (TDB) \\
Orbital eccentricity, $e$ & $< 0.03$ \\
Pulsar mass function, $f_x$ ($10^{-8} M_{\odot}$) & 9.12(2) \\
Minimum donor mass, $m_d$ ($M_{\odot}$) & 0.005 - 0.007 \\
Maximum $Z_1^2$ power & 355.4 \\
\tableline
\end{tabular}}
\tablecomments{Parameter uncertainties for $\nu_0$, $a_x\sin(i)/c$,
  $P_{orb}$ and $T_0$ are $3\sigma$ ($\Delta Z_1^2 = 9$) values, given
  in the last quoted digits.  Limits for $e$, and $\dot\nu$ are
  $1\sigma$. The minimum donor mass range is for neutron star masses
  of 1.2 and 2 $M_{\odot}$.}
\end{table*}

\section{Discussion and Summary}

Our analysis of observations of \source{} with {\it NICER} obtained in
2017 August, October and November confirms the discovery by SK17 that
it is a 163.656 Hz AMXP, and allowed us to derive the orbital
parameters of the system for the first time.  The 37.97 min orbital
period of \source{} is the shortest currently known for an AMXP, and
our measurement confirms several previous indirect indications that
the system is an ultracompact binary \citep{2018arXiv180103006H,
  2018MNRAS.475.2027V}. We measure a mass function, $f_x = (m_d
\sin i)^3 / (m_{ns} + m_d)^2 = ( (a_x\sin i)^3 \omega_{orb}^2 )/G =
9.12 \times 10^{-8} M_{\odot}$, which is also the smallest among
stellar binaries.  The mass function defines a lower limit to the mass
of the donor star, $m_d$. For a neutron star mass in the range from
$m_{ns} = 1.2 - 2 M_{\odot}$ we find a minimum donor mass in the range
from $0.005 - 0.007 M_{\odot}$.  Given the orbital period and a
plausible range of total system mass, the separation between the
components is of order $300,000$ km, and would fit within the Earth -
Moon distance.

The reasonable assumption that the donor star fills its Roche lobe
provides a constraint on the mean density of the donor.  This can be
expressed as a constraint on its radius in units of the component
separation, $a$, that depends principally on the system's mass ratio,
$q = m_d / m_{ns}$ \citep{1983ApJ...268..368E}. We combine this
constraint with that from the measured mass function to explore the
implications for the nature of the donor star and the system's orbital
inclination.

Our results are summarized in Figure 7 which shows constraints on the
donor mass and radius.  We plot the Roche lobe constraint for three
different neutron star masses, $1.2$ (green), $1.4$ (black), and $1.8$
$M_{\odot}$ (red). The closeness of the three curves is a visual
demonstration of how insensitive this constraint is to the assumed
neutron star mass.  The different symbols along the curve mark
inferred donor masses from the mass function constraint for different
assumed orbital inclinations, $i$, and for two values of the neutron
star mass at each inclination. For each pair of symbols the left- and
right-most correspond to a neutron star mass of $1.2$ and $1.8
M_{\odot}$, respectively.  The left-most symbol for $i = 90^{\circ}$
marks the minimum donor mass for a $1.2 M_{\odot}$ neutron star.
First, we note that the constraints require hydrogen-deficient donors,
as is expected for systems with an orbital period less than about 80
min \citep{1984ApJ...283..232R, 2002ApJ...577L..27B}. For additional
context we show mass - radius relations obtained from the literature
for several donor types.  The dashed curve is the mass - radius
relation for low mass, cold, pure helium white dwarfs from
\cite{1969ApJ...158..809Z}, as corrected by
\cite{1984ApJ...283..232R}. Here we have plotted it using the fitting
formula of \cite{2001A&A...368..939N}. The dotted curves denote a
range of mass - radius values from the binary evolutionary
calculations of \citet{2007MNRAS.381..525D} for the helium donors of
AM CVn systems.  The region between the upper and lower dotted curves
gives an indication of the allowed range in mass and radius for donors
at different evolutionary stages, and with different values of central
degeneracy at the onset of mass transfer (see Deloye et al. 2007 for
details). Lastly, the dash-dotted curves show mass - radius relations
for carbon white dwarfs with central temperatures of $10^4$ (lower)
and $3 \times 10^6$ K (upper), from \citet{2003ApJ...598.1217D}. Thus,
\source{} appears to be a somewhat more extreme example of the
currently known ultracompact AMXPs \citep{2007ApJ...668L.147K,
  2012arXiv1206.2727P}.

For a random distribution of inclination angles the chance probability
to observe a system with an inclination less than or equal to $i$ is
$1 - \cos(i)$. The probability to observe an inclination angle less
than or equal to $\approx 18.2^{\circ}$ is $5\%$.  From this, and
assuming a $1.8 M_{\odot}$ neutron star mass, we deduce a $95\%$
confidence upper limit to the donor mass of $0.0216 M_{\odot}$, with a
corresponding radius of $0.05 R_{\odot}$, substantially less than that
of any hydrogen-rich brown dwarfs \citep{2000ApJ...542..464C}.  We use
this upper limit on the donor radius to place an upper limit on the
inclination of $\approx 84^{\circ}$, since no eclipses are seen in the
light curve.

Additional insight is provided by estimates of the long term mass
accretion rate, $\dot M$, and the realization that the mass transfer
in such systems is driven by angular momentum loss due to
gravitational radiation \citep{1984ApJ...283..232R,
  2001ApJ...557..292B}.  Interestingly, for \source{} we have $\dot M$
estimates from both persistent X-ray flux measurements and modeling of
its thermonuclear X-ray bursts that are in substantial agreement
\citep{2017ApJ...836..111K}, and suggest $\dot M \approx 2.5 \times
10^{-11}$ $M_{\odot}$ yr$^{-1}$.  This value for \source{} is also in
general agreement with the calculations of $\dot M$ versus $P_{orb}$
reported by Deloye et al. (2007, see their Figure 15).  Based on this,
and the reasonable assumption that the donor responds to mass loss
like a degenerate star \citep{2001ApJ...557..292B}, we can estimate
the donor mass as $m_d = 0.0175 (m_{ns}/ 1.4 M_{\odot})^{-1/3} \;
M_{\odot}$.  Using this result we additionally show in Figure 7 (with
blue ``$+$'' symbols) the donor masses and corresponding radii for
neutron star masses of 1.4 and 2 $M_{\odot}$, where the higher donor
mass estimate corresponds to the lower mass neutron star (1.4
$M_{\odot}$).  These mass estimates further imply constraints on the
binary inclination angle of $19^{\circ} < i < 27.5^{\circ}$, where the
lower and upper bounds correspond to neutron star masses of 1.4 and 2
$M_{\odot}$, respectively.

The constraints summarized in Figure 7 suggest that \source{} is
observed at relatively low inclination, and the donor mass - radius
constraints appear to be consistent with the helium donors of AM CVn
systems explored by \citep{2007MNRAS.381..525D} (the dotted curves in
Figure 7).  We note that these authors also provide estimates of the
expected orbital period evolution, $\dot P_{orb}$, for these systems.
Given the observed orbital period of \source{}, the predicted values
are in the range $\dot P_{orb} \approx 1 - 3 \times 10^{-6}$ s
yr$^{-1}$, which can be probed with additional {\it NICER} timing
observations.

Clues to the donor composition in a neutron star X-ray binary can in
principle be provided by the properties of its thermonuclear flashes.
The energetic, long duration burst events seen to date would appear to
be consistent with deep ignition of a helium-rich layer
\citep{2017ApJ...836..111K}. As noted previously, under certain
conditions the stable burning of accreted hydrogen into helium can
result in helium-powered thermonuclear flashes
\citep{1981ApJ...247..267F, 2006ApJ...652..559G}. Our measurements
confirm the previous indications for a hydrogen-deficient donor in
\source{} \citep{2018arXiv180103006H, 2018MNRAS.475.2027V}, and
definitively rule out this option, since the accreted fuel cannot
contain a significant fraction of hydrogen.

%and thus a significant helium fraction in
%the donor.  This conclusion is strengthened by our requirement for a
%hydrogen-deficient donor in \source{}. Previously one could appeal to
%burning of accreted hydrogen as a source for the helium fuel, but our
%measurements now rule out this option, since the accreted fuel cannot
%contain a significant fraction of hydrogen. 
%We note that some
%indications of a significantly truncated accretion disk in \source{}
%based on X-ray spectroscopy have been reached under the assumption
%that the inclination angle was relatively high
%\citep{2017MNRAS.464..398D, 2018MNRAS.475.2027V}. Indeed, van den
%Eijnden et al. (2018) commented that at low inclination a disk
%extending down to the neutron star could not be ruled out.
%We suggest that such
% spectroscopic conclusions relying on a high inclination be reassessed
% in light of the evidence above suggesting a low to moderate orbital
% inclination for \source{}.

While a helium donor in \source{} appears quite plausible given the
measurements presented here, as well as its bursting properties, prior
X-ray spectroscopy results have suggested \source{} may have an
oxygen-rich circumbinary environment, perhaps associated with an
outflow \citep{2018MNRAS.475.2027V}. In addition, spectral modeling of
the Fe K$\alpha$ reflection feature appears to favor a higher
inclination than suggested by our constraint derived from the
assumption of gravitational radiation driven mass loss
\citep{2017MNRAS.464..398D, 2018MNRAS.475.2027V, 2017ApJ...836..111K}.
Based on these indications, \cite{2018MNRAS.475.2027V} favor a CO or
O-Ne-Mg white dwarf donor. Given the constraints on the donor
summarized in Figure 7 this remains a viable option, particularly in
the case of non-conservative mass transfer, as would occur in the
presence of an outflow.  However, such a conclusion would also open up
additional questions, such as the nature of the fuel for the observed
X-ray bursts, which is presumably helium \citep{2017ApJ...836..111K},
though we note that \cite{2018arXiv180103006H} did not detect helium
in their optical spectra of \source{}.  Further to this final point,
the bursting low mass X-ray binary 4U 0614$+$091 is another source
with apparently helium-powered X-ray bursts
\citep{2010A&A...514A..65K}, but whose optical spectra are suggestive
of a CO donor with little to no helium \citep{2006A&A...450..725W,
  2004MNRAS.348L...7N}. Additional observations will likely be needed
to definitively pin down the nature of the donor in \source{}.

 While most AMXPs are transient systems, \source{} is distinctive in
 that it has been in outburst now for about a decade.  This provides
 an exciting opportunity to study the long-term spin and orbital
 evolution with additional {\it NICER} observations.  Moreover, we now
 have detections of pulsations from \source{} at two widely spaced
 epochs, in 2008 May with {\it RXTE}, and the present 2017 August,
 October and November observations with {\it NICER}.  Interestingly,
 the source shows some indications of a significant change in pulsed
 amplitude in that time-frame.  The estimated source pulsed amplitude
 measured by SK17 with {\it RXTE} was $9.4 \pm 1.1\%$ ($2 - 12$ keV),
 whereas we find $2.04 \pm 0.11 \%$ ($0.3 - 3.2$ keV) with {\it
   NICER}.  We note that given the current uncertainties associated
 with modeling the {\it NICER} background, combined with the fact that
 the source count rate is dropping steadily above $\approx 5$ keV, it
 is presently challenging to accurately determine the pulsed amplitude
 above this energy. Nevertheless, with the present data we can measure
 the pulsed amplitude in the $2 - 5$ keV band with reasonable
 precision, and we find a value of $3.2 \pm 0.3 \%$.  Based on this we
 think it likely that the smaller amplitude measured by {\it NICER} is
 a real effect and likely represents some secular change within the
 system, perhaps associated with the effect of accretion on the
 magnetic field, as, for example, suggested by
 \cite{2012ApJ...753L..12P}.  More definitive conclusions in this
 regard should become feasible as the {\it NICER} background
 calibration improves.  We will pursue this, as well as searches for
 energy dependent phase lags and a detailed spectroscopic study in
 subsequent work.

%Such a clear drop in amplitude could account for the
%``intermittency'' of pulsations seen in some systems, as this
%relatively low amplitude could be difficult to detect with less
%sensitive timing instruments.  Additional {\it NICER} observations
%should be able to track any further changes in pulsed amplitude.

\label{sec:conclusions}

\acknowledgments

This work was supported by NASA through the {\it NICER} mission and
the Astrophysics Explorers Program. This research also made use of
data and/or software provided by the High Energy Astrophysics Science
Archive Research Center (HEASARC), which is a service of the
Astrophysics Science Division at NASA/GSFC and the High Energy
Astrophysics Division of the Smithsonian Astrophysical Observatory. SG
acknowledges the support of the Centre National d’Etudes Spatiales
(CNES).  We thank the anonymous referee for a helpful report.

\facility{NICER, ADS, HEASARC}

\software{HEAsoft (v6.22; Arnaud 1996), mpfit (Markwardt 2009)}

%LK thanks the International Space Science Institute in
%Bern, Switzerland and the National Science Foundation under Grant
%No. PHY-1430152 (JINA Center for the Evolution of the Elements) for
%supporting events that benefited this work. We thank the anonymous
%referee for a helpful review.

\bibliographystyle{aasjournal}

\bibliography{ms}

\newpage

% Maximum 5 Figures + Tables for ApJ Letter

\begin{figure*}
\begin{center}
\includegraphics[scale=0.75]{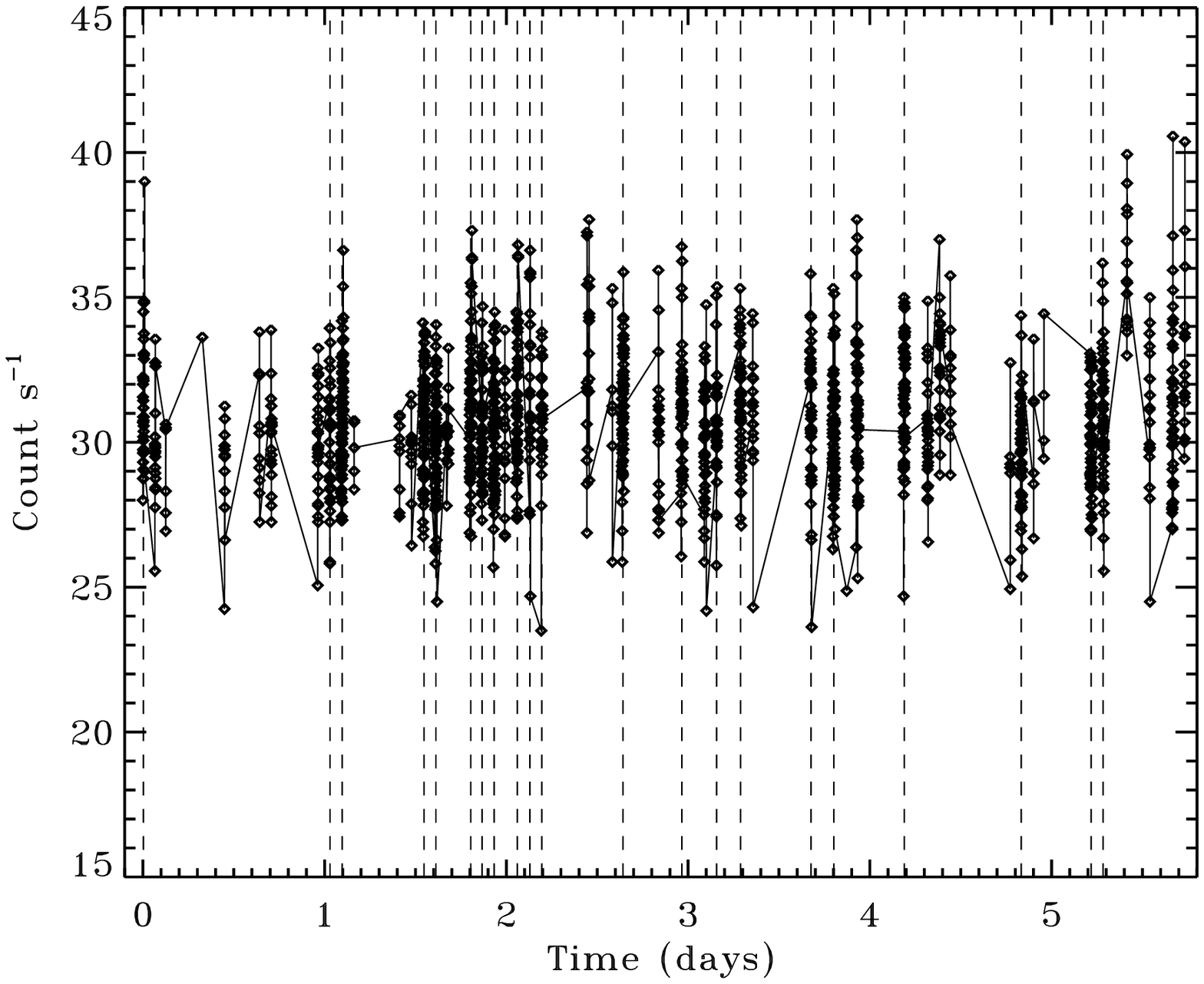}%
\end{center}
\caption{\label{fig:lc} Light curve of \source{} from {\it NICER}
  observations obtained in 2017 August. Data are the summed counting
  rates in 16 s bins in the 0.3 - 5 keV band. The vertical dashed
  lines mark the centers of 21 good time intervals used to compute
  average power spectra (see \S 2). Time zero is MJD 57974.8334963496
  (TDB). }
\end{figure*}

%\vfill\eject
%\newpage

\begin{figure*}
\begin{center}
\includegraphics[scale=0.75]{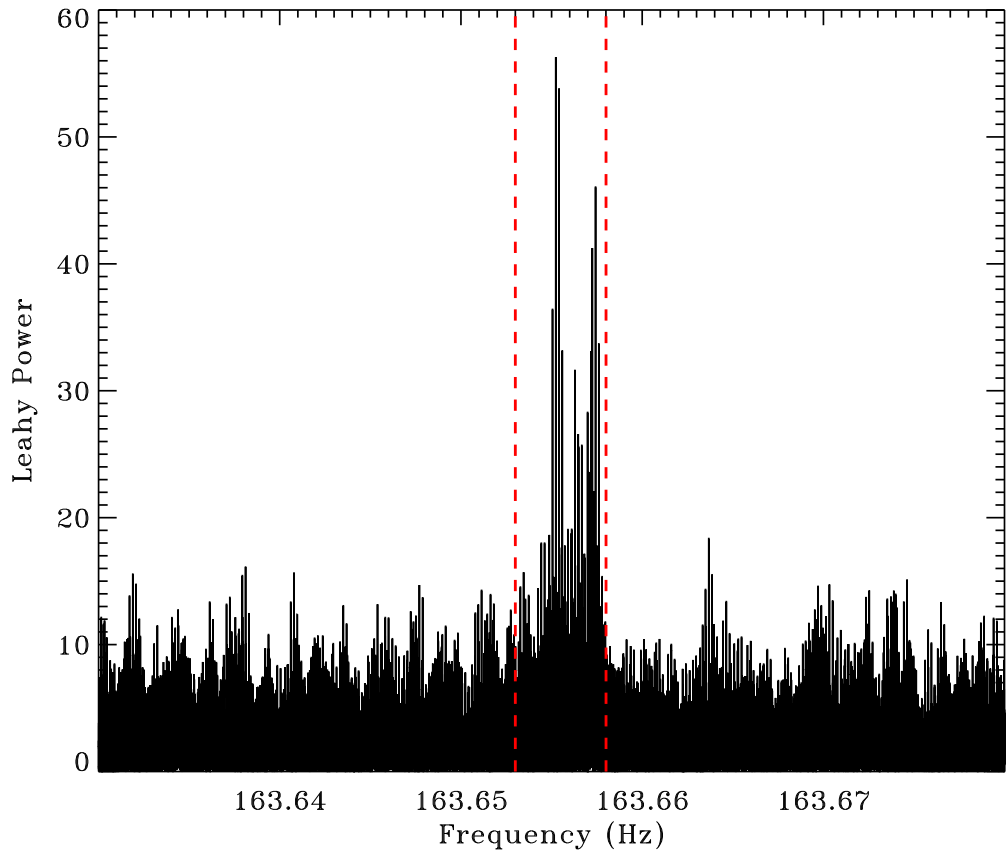}%
\end{center}
\caption{\label{fig:discover_pds} Power spectrum of \source{} from
  {\it NICER} observations obtained in 2017 August, in the vicinity of
  the pulse frequency detected with {\it RXTE} by SK17. The spectrum
  was computed from a light curve spanning 500 ksec and sampled at
  4096 Hz, and includes events in the 0.3 - 3.2 keV band. The 163.656
  Hz pulsar peak is clearly evident. The vertical red lines indicate
  the approximate range of pulse frequency detected in the {\it RXTE}
  observations. See \S 2 for a detailed discussion of the pulsation
  search. }

\end{figure*}

%\vfill\eject
%\newpage

\begin{figure*}
\begin{center}
\includegraphics[scale=0.8]{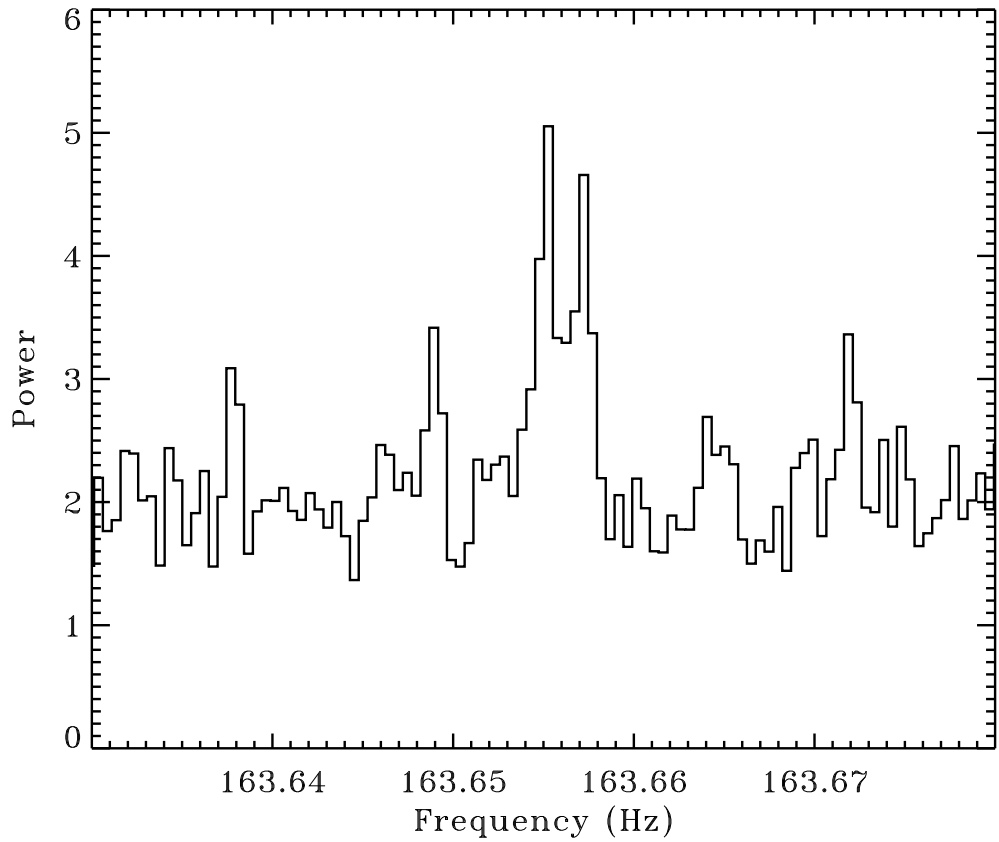}%
\end{center}
\caption{\label{fig:avg_pds_dwells} Average power spectrum of
  \source{} from {\it NICER} observations obtained in 2017 August, in
  the vicinity of the 163.656 Hz pulsar frequency, computed from 21
  on-source dwells with exposures ranging from 1294 s (longest) to 545
  s (shortest).  The pulsar signal is comprised of two dominant
  side-bands. See \S 2 for a detailed discussion of the pulsation
  search. }

\end{figure*}

%\vfill\eject
%\newpage

\begin{figure*}
\begin{center}
\includegraphics[scale=0.75]{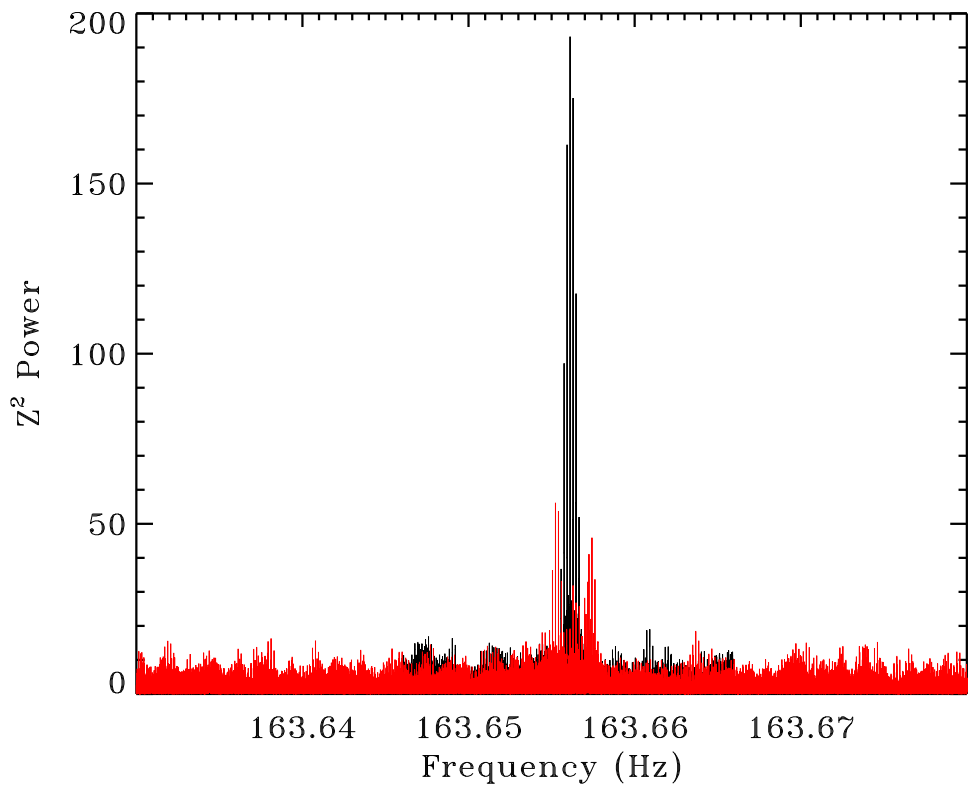}%
\end{center}
\caption{\label{fig:z2_compare} Comparison of the $Z_1^2$ signals for
  the August {\it NICER} data with and without the orbital phase
  delays. The curves show $Z_1^2$ evaluated on a grid of pulsar
  frequency, $\nu_0$, with orbit phase delays included (black), and
  without (red). The orbit solution recovers a single, coherent peak
  with $Z_1^2 = 196.1$, modulated by the window function imposed by
  visibility from the {\it ISS} orbit. See \S 3 for a detailed
  discussion of the orbit search. }
\end{figure*}

\begin{figure*}
\begin{center}
\includegraphics[scale=0.75]{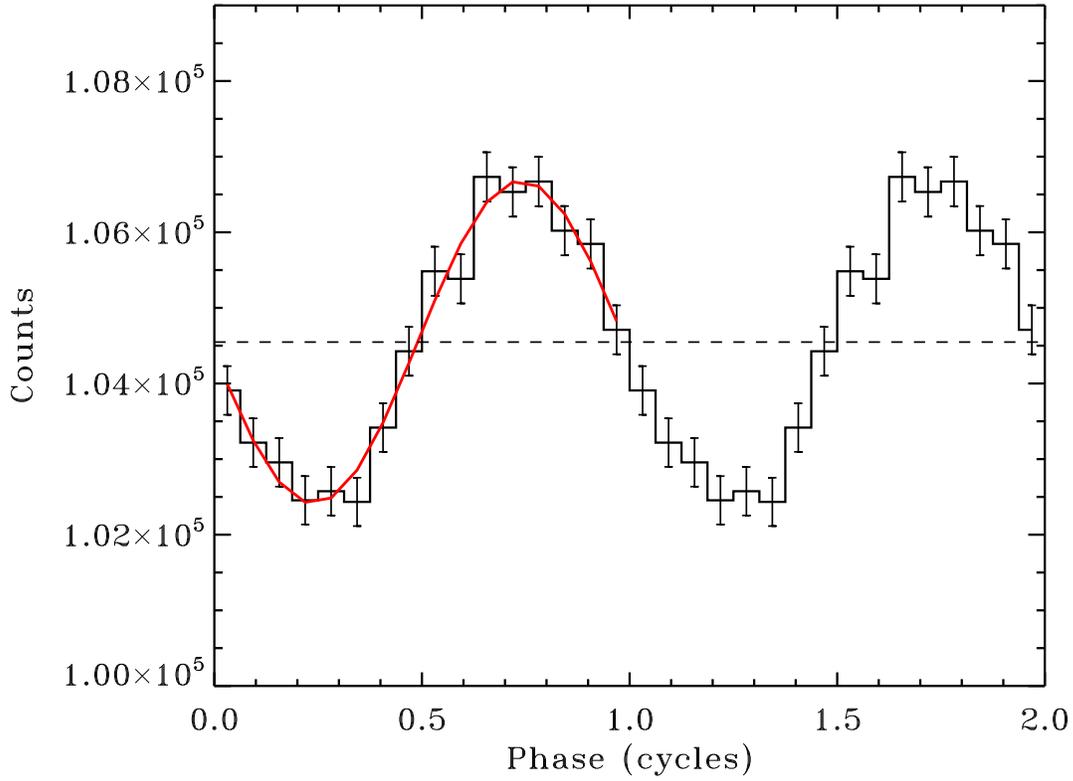}%
\end{center}
\caption{\label{fig:pulse_profile} Pulse profile obtained after phase
  folding the August, October and November {\it NICER} exposures of
  \source{} with the best-determined orbital solution.  The profile
  includes events in the 0.3 - 3.2 keV range, and we used 16 phase
  bins.  The best fitting sinusoid, $A + B\sin(\phi)$, is also plotted
  (red).  The fit has $\chi^2 = 8.6$ with 13 degrees of freedom. The
  pulsed amplitude is $B/A = 2.04 \pm 0.11 \%$.  }
\end{figure*}

\begin{figure*}
\begin{center}
\includegraphics[scale=0.75]{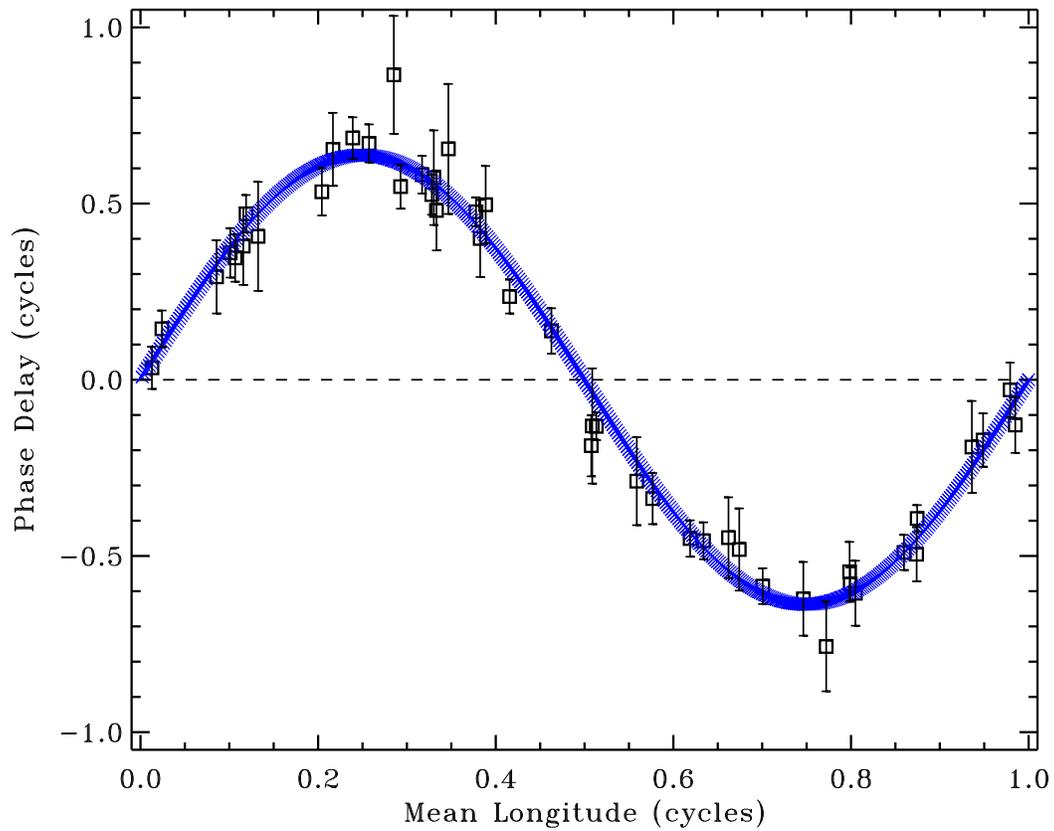}%
\end{center}
\caption{\label{fig:phase_delays} Pulse phase residuals from {\it
    NICER} observations of \source{} as a function of mean pulsar
  longitude computed using the best orbit solution. The residual is
  plotted in units of pulsar phase (cycles), and the orbit-predicted
  phase delay is added to the residuals to show the orbital
  variations.  }
\end{figure*}

\begin{figure*}
\begin{center}
\includegraphics[scale=0.75]{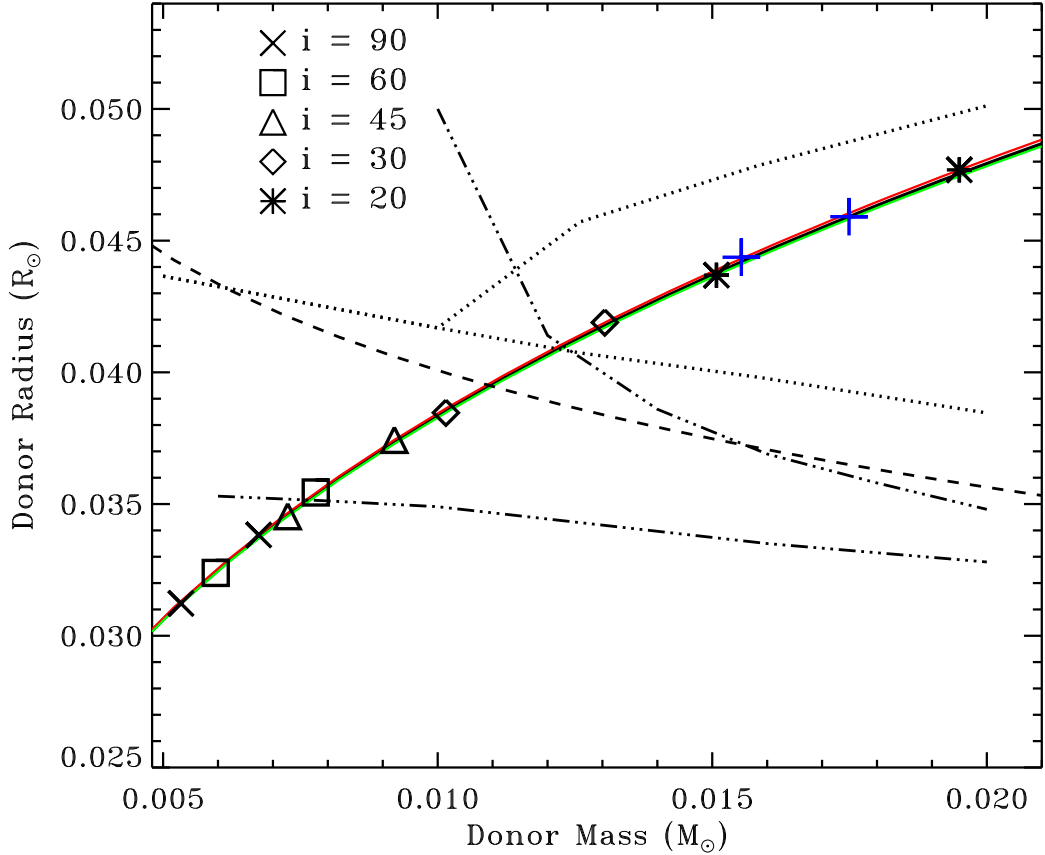}%
\end{center}
\caption{\label{fig:mrdonor} Constraints on the donor in
  \source{}. The Roche lobe constraint is plotted for three different
  neutron star masses, $1.2$ (green), $1.4$ (black), and $1.8$
  $M_{\odot}$ (red). The different symbols along the curves denote
  donor masses from the mass function constraint for different assumed
  inclinations, $i$, and for two values of the neutron star mass at
  each inclination. For each pair of symbols the left- and right-most
  correspond to neutron stars of $1.2$ and $1.8 M_{\odot}$,
  respectively.  Also shown are mass - radius relations obtained from
  the literature for several donor types.  The dashed curve is the
  fitting formula from \cite{2001A&A...368..939N} that approximates
  the mass - radius relation for low mass, cold, pure helium white
  dwarfs \citep{1969ApJ...158..809Z}. The dotted curves denote a range
  of mass - radius values from the binary evolutionary calculations of
  \citet{2007MNRAS.381..525D} for the helium donors of AM CVn
  systems. The dash-dotted curves show mass - radius relations for
  carbon white dwarfs with central temperatures of $10^4$ (lower) and
  $3 \times 10^6$ K (upper), from \citet{2003ApJ...598.1217D}.
  Lastly, the blue ``$+$'' symbols mark the masses and radii that
  would produce a long-term $\dot M = 2.5\times 10^{-11} M_{\odot}$
  yr$^{-1}$ via gravitational radiation for neutron stars of 1.4
  (higher value) and 2 $M_{\odot}$ (lower value, see \S 4 for further
  discussion). }
\end{figure*}


\end{document}